\documentclass[aps,pra,superscriptaddress,amsmath,amssymb,amsfonts,twocolumn,nofootinbib,longbibliography]{revtex4-2}

\usepackage{amssymb}
\usepackage{dcolumn}
\usepackage{graphicx}
\usepackage{bbm}
\usepackage{amssymb,amsmath}
\usepackage{amsfonts}
\usepackage{hyperref}
\usepackage{color}
\usepackage{soul,xcolor}
\usepackage[normalem]{ulem}
\usepackage{bm}
\usepackage{braket}
\usepackage{amsmath}
\usepackage{graphicx,color,xcolor,colortbl}
\newcommand{\be}{\begin{equation}}
\newcommand{\ee}{\end{equation}}
\newcommand{\beq}{\begin{eqnarray}}
\newcommand{\eeq}{\end{eqnarray}}

\newcommand{\hide}[1]{}

\begin{document}

\title{Breaking and trapping Cooper pairs by Rydberg-molecule spectroscopy\\ in atomic Fermi superfluids}

\author{Chih-Chun Chien}
\email{cchien5@ucmerced.edu}
\affiliation{ITAMP, Center for Astrophysics $|$ Harvard $\&$ Smithsonian, Cambridge, USA}
\affiliation{Department of Physics, University of California, Merced, CA, USA}

\author{S.~I.~Mistakidis}
\affiliation{ITAMP, Center for Astrophysics $|$ Harvard $\&$ Smithsonian, Cambridge, USA}
\affiliation{Department of Physics, Missouri University of Science and Technology, Rolla, MO, USA}

\author{H.~R.~Sadeghpour}
\affiliation{ITAMP, Center for Astrophysics $|$ Harvard $\&$ Smithsonian, Cambridge, USA}
\begin{abstract}
We propose a spectroscopic probe of the breaking and localization of Cooper pairs in an atomic Fermi superfluid interacting with a Rydberg impurity. This is achieved by monitoring the formation of diatomic and triatomic ultralong-range molecular species in the superfluid across the Bardeen–Cooper–Schrieffer (BCS)-Bose Einstein condensation (BEC) crossover. 
The triatomic Rydberg molecule in the BEC regime heralds the trapping of a tightly-bound Cooper pair, reminiscent of pion capture in nuclear matter, while the breaking of a Cooper pair on the BCS side by a diatomic Rydberg molecule is evocative of binary-star tidal disruption by a black hole. Spectroscopy of the Fermi superfluid and Rydberg molecules allows for an estimation of the Cooper-pair size while the Rydberg molecule binding energies discern many-body pairing effects.
\end{abstract}
\maketitle

Rydberg atom-based systems have emerged as leading platforms for demonstrating many-body correlations~\cite{Antoine20}, quantum simulations~\cite{RevModPhys.82.2313,Adams_2020,Wu_2021}, quantum error corrections~\cite{Bluvstein24}, and quantum optics~\cite{Shao24}. When the excited electron scatters from a nearby ground-state atom, under certain conditions, ultralong-range molecular bonds can form~\cite{PhysRevLett.85.2458,Chibisov02,Hamilton02}.
Such long-range Rydberg molecules have been realized ~\cite{Bendkowsky2009,Niederprum2016,Althon2023,Booth15,PhysRevA.101.060701,Shaffer2018,Fey20}. Interesting aspects of many-body physics, such as the formation of Bose and Fermi polarons, quantum statistics of gases exhibiting bunching and anti-bunching, with Rydberg molecules have also been reported~\cite{PhysRevLett.120.083401,PhysRevLett.116.105302,PhysRevA.97.022707,Sous20,Durst24}. These studies exploit the large energy separations between the vibrational energies and the underlying primitive excitations in a quantum gas.

In a different context, two-component fermions with attractive interactions form Cooper pairs and exhibit the Bardeen-Cooper-Schrieffer (BCS) - Bose-Einstein condensation (BEC) crossover pioneered by the experiments with ultracold Fermi gases~\cite{crossover,PhysRevLett.92.120403,PhysRevLett.92.203201} (see also  \cite{Pethick-BEC,Ueda-book,ZwergerBook}). 
Here, we show that by creating ultralong-range molecules with a Rydberg impurity in a background sea of Cooper pairs, it is possible to a) break the pairs on the BCS side and b) locally trap a Cooper pair on the BEC side. 
The former bears analogies with the 
breaking of a binary-star pair by a tidal disruption event into a black hole~\cite{Hill98,Bromley_2012}, while the latter 
is reminiscent of the capture of pions (quark-antiquark pairs) in hydrogen~\cite{Gotta2008,Hirtl2021}, deuterium~\cite{Strauch2011}, and helium~\cite{SciPostPhysProc.5.026}.

By radio-frequency (rf) spectroscopy of the superfluid pairing gap~\cite{PhysRevLett.99.090403,doi:10.1126/science.aan5950} or Rydberg spectroscopy of the molecular lines~\cite{PhysRevLett.120.083401,PhysRevA.97.022707}, one may, in a local spectroscopic manner, probe the reaction of the superfluid to tackle topical problems in condensed matter physics, such as 
the Cooper-pair size and pairing energies~\cite{Schunck2008Nature}.

\begin{figure}[t]
\centering
\includegraphics[width=\columnwidth]{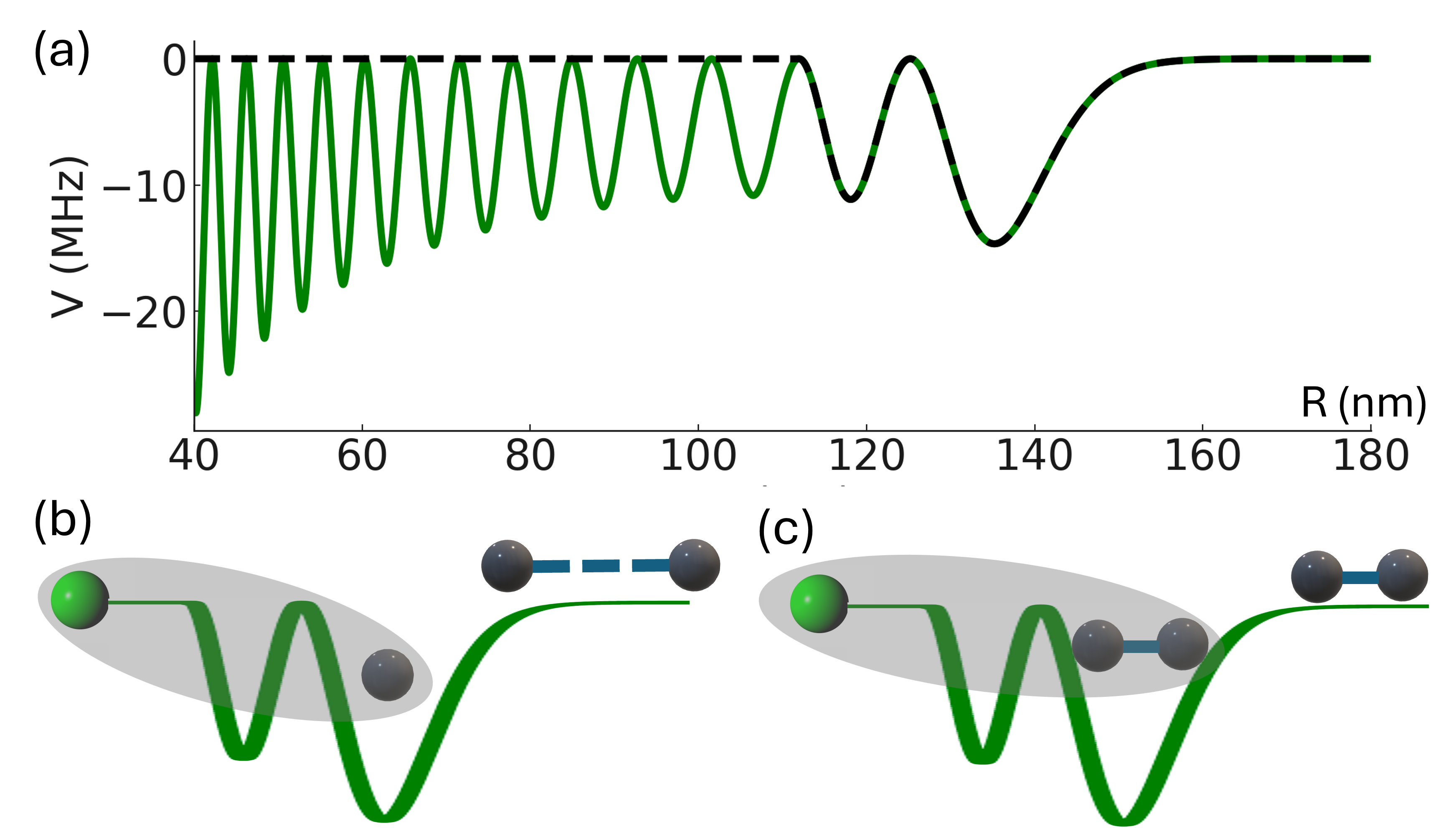}
\caption{(a) A typical Rydberg potential (solid line) and the outer double-well approximation (dashed line). The formation of (b) a diatomic Rydberg molecule with a fermion from a broken Cooper pair and (c) a triatomic Rydberg molecule with a Cooper pair. The green (black) spheres represent the Rydberg atoms (fermions in the superfluid). The Cooper pairs are visualized by two black spheres connected by a dashed or solid line. The grey ellipsoids mark the Rydberg molecules.}
\label{Fig:Demo}
\end{figure}

A typical Rydberg potential is illustrated in Fig.~\ref{Fig:Demo} with 
heteronuclear Rydberg molecules formed in Fermi superfluids.
These molecules form (1) in a diatomic bond between the impurity and a fermion from a broken Cooper pair and (2) in a triatomic bond - a Cooper pair in a molecule - with the pair trapped in the Rydberg potential. To our knowledge, the latter type of pair trapping has not previously been discussed, and these molecules are different from the trimer Rydberg molecules emanating from two weakly interacting bosons individually trapped by a bosonic Rydberg atom that have been realized~\cite{PhysRevLett.105.163201} and theoretically studied~\cite{PhysRevLett.102.173001,Fernandez16}. This work exploits the interplay between two molecule-formation mechanisms, one between the fermions to bind a Cooper pair and the other among the Rydberg and its neighboring atoms to create a Rydberg molecule. Similar competitions influence the physics of the aforementioned astro and nuclear physics examples.

Leveraging the Bogoliubov-de Gennes (BdG) formalism~\cite{degennes-sc,BdG-book} suitable for studying inhomogeneous effects in Fermi superfluids, we extract the low-lying bound states of the composite Rydberg-Fermi superfluid system. By increasing the Cooper-pair strength, distinct local reactions of the pairing gap will occur: Breaking (trapping) of a weak (strong) Cooper pair leads to a local suppression (enhancement) of the gap function. We identify the formation of diatomic and triatomic Rydberg molecules along with their binding energies, which are raised by the many-body pairing effect when compared to those in a noninteracting gas. 
In contrast to previous studies~\cite{PhysRevLett.110.020401} with impurities carrying onsite potentials in Fermi superfluids, the Rydberg potential has its furthest well about hundreds of nanometers away from the core with controllable width and depth, thereby giving rise to rich structures of Rydberg molecules.

\paragraph*{\textit{Rydberg excitation in a Fermi superfluid}.--} 
We consider few Rydberg atoms immersed in  
a two-component, spin- and mass-balanced Fermi superfluid with contact pairing interactions. Experimentally, this setup can be emulated, for instance, by bosonic $^{87}$Rb Rydberg atoms and two hyperfine states of $^{84}$Rb or $^{86}$Rb for the Fermi superfluid. Relevant experimental progress towards such atomic mixtures can be found in Ref.~\cite{PhysRevA.62.011402}. However, we emphasize that our results equally hold for other Rydberg atom-Fermi superfluid systems.
For simplicity, the Rydberg atoms are assumed to be immobile and noninteracting with each other. 
A quasi one-dimensional (1D) geometry~\cite{MISTAKIDIS20231} creating a cigar-shaped cloud similarly to  Refs.~\cite{PhysRevLett.92.203201,PhysRevLett.117.235301} is considered. It supports the off-diagonal long-range order of the superfluid while freezing out the transverse degrees-of-freedom. 

The many-body Hamiltonian of the composite system within the BCS-Leggett theory reads 
\begin{eqnarray}\label{Eq:Hint}
\mathcal{H}=\mathcal{H}_{BCS}+\sum_{\sigma}\int dx V_{Ryd}(x)\psi_\sigma^{\dagger}(x)\psi_\sigma(x) d^\dagger d, 
\end{eqnarray}
where $\mathcal{H}_{BCS}=\int dx\Big[\sum_{\sigma} \psi_{\sigma}^{\dagger}(x)h_{\sigma}(x)\psi_{\sigma}(x)+ (\Delta(x) \psi_{\uparrow}^{\dagger}(x)\psi_{\downarrow}^{\dagger}(x)+h.c.)\Big]$
\cite{Leggett,Pethick-BEC}. 
The fermion operator acting on the $\sigma=\uparrow,\downarrow$ component of mass $m$ is $\psi_\sigma$, and $h_\sigma(x)=-\frac{\hbar^2}{2m}\nabla^2+V_{ext}(x)-\mu_\sigma$ denotes the single-particle Hamiltonian with $V_{ext}(x)$ summarizing the total external confinement. 
The order parameter of the $s$-wave Fermi superfluid is the gap function  
\begin{align}\label{Eq:DeltaOriginal}
\Delta(x)&=-U\langle\psi_{\downarrow}(x)\psi_{\uparrow}(x)\rangle.
\end{align}
The effective coupling $U<0$ is related to the 1D scattering length $a_{1D}$~\cite{Olshani} via $U=-\frac{2\hbar^2}{ma_{1D}}$, tunable by Feshbach resonance~\cite{Pethick-BEC}, and 
$\langle \dots \rangle$ designates the ground-state expectation value at $T=0$. 
The BCS-BEC crossover occurs when the chemical potential (here $\mu_{\uparrow}=\mu_{\downarrow} \equiv \mu$) crosses zero~\cite{ZwergerBook} where the minimum of the quasiparticle-spectrum shifts to zero momentum.  

Importantly, the second contribution in Eq.~\eqref{Eq:Hint} models the Rydberg atom - fermion interaction~\cite{Sous20}, with $d$ ($d^{\dagger}$) being the annihilation (creation) operator of a Rydberg atom. The ultralong-range Born-Oppenheimer potential between a Rydberg atom and a ground-state fermionic atom is given by 
$V_{Ryd}(x)=\frac{2\pi\hbar^2 a_e}{m_e}|\psi_e(x)|^2$ \cite{PhysRevLett.85.2458}. Here, $a_e$ denotes the scattering length between the Rydberg electron with mass $m_e$ and a fermionic atom, and $x$ measures the distance from the Rydberg impurity to the fermionic atom. %
The Rydberg electron wave function, $\psi_e(x)$, is calculated with effective valence potentials~\cite{marinescu94}.
In the vicinity of a Rydberg atom, we replace $d^\dagger d$ by $\langle d^\dagger d\rangle =1$ and hence $V_{Ryd}(x)$ acts as 
an effective potential for the Fermi superfluid. In what follows, the localized Rydberg potential will be implicitly combined with $V_{ext}(x)$ in $h_\sigma$. Moreover, the Rydberg potential is approximated by the double-well form shown in Fig.~\ref{Fig:Demo}(a) since the outer two wells represent the two largest lobes of the Rydberg electron wave function of interest (see Ref.~\cite{mixing_note} for more information), and therefore have the largest Frank-Condon factor for excitation. 
The Fermi energy $E_f=\hbar^2 k_f^2/(2m)$ and wavevector $k_f=\pi n/2$, of a noninteracting 1D two-component Fermi gas with the same total particle number $N=\int n(x) dx $ as the superfluid serve as the energy and inverse-length units. For example, $g=-U k_f/E_f$ is the dimensionless pairing strength. Here, mean-field theory is used to describe ground-state properties of the quasi-1D system. If critical behavior  
is encountered, more 
sophisticated theories may be consulted~\cite{ZwergerBook}.

\paragraph*{\textit{ BdG formalism.} --} To reveal the impact of the Rydberg atoms on the Fermi superfluid, we inspect the composite system as the superfluid undergoes the BCS-BEC crossover. Specifically, $\mathcal{H}$ can be diagonalized by the BdG transformation \cite{bogoliubov1947theory,BdG-book}: 
     $\psi_{\uparrow,\downarrow}(x)= \sum_{\tilde{n}}[u_{\uparrow,\downarrow}^{\tilde{n}1,2}(x)\gamma_{\tilde{n}1,2}\mp v_{\uparrow,\downarrow}^{\tilde{n}2,1*}(x)\gamma_{\tilde{n}2,1}^{\dagger}]$. 
The quasiparticle wave functions  $u_\sigma^{\tilde{n}j}$ and $v_\sigma^{\tilde{n}j}$ with $j=1,2$ are to be determined, and they  satisfy $\int dx(|u_\sigma^{\tilde{n}j}|^2+|v_\sigma^{\tilde{n}j}|^2)=1$. 
The BdG equation for the composite system considered here can be block-diagonalized into~\cite{BdG-book} 
\begin{equation}\label{eq:subset1}
  \begin{pmatrix}
    h_{\uparrow}(x)&\Delta(x)\\
    \Delta^*(x)&-h^*_{\downarrow}(x)
    \end{pmatrix}
    \begin{pmatrix}
    u^{\tilde{n}j}_{\uparrow}(x)\\
    v^{\tilde{n}j}_{\downarrow}(x)
    \end{pmatrix}
    =E_{\tilde{n}j}\begin{pmatrix}
    u^{\tilde{n}j}_{\uparrow}(x)\\
    v^{\tilde{n}j}_{\downarrow}(x)
    \end{pmatrix}.
\end{equation}
Moreover, the BdG equation has a discrete symmetry connecting the positive and negative energy states, 
so we drop the indices $1,2$ and $\uparrow,\downarrow$ from the quasi-particle wave functions. 
For the ground state, the gap function Eq.~(\ref{Eq:DeltaOriginal}) then becomes
$\Delta(x)=-U{\sum_{\tilde{n}}}' u^{\tilde{n}}_{\uparrow }(x)v^{\tilde{n}*}_{\downarrow}(x)$ and the total fermion density $n(x)=\sum_\sigma n_{\sigma}(x)= 
2{\sum_{\tilde n }}'|v_{\tilde n}(x)|^2$ with      $n_\sigma(x)=\langle\psi_\sigma^\dagger(x)\psi_\sigma(x)\rangle$. Here ${\sum_{\tilde{n}}}'$ denotes summation over the positive-energy states. We discretize the space and implement an iterative method~\cite{BdG-book,PhysRevA.107.063314} to solve the BdG equation (see SM~\cite{supp} for details).

\begin{figure}[t]
\centering
\includegraphics[width=\columnwidth]{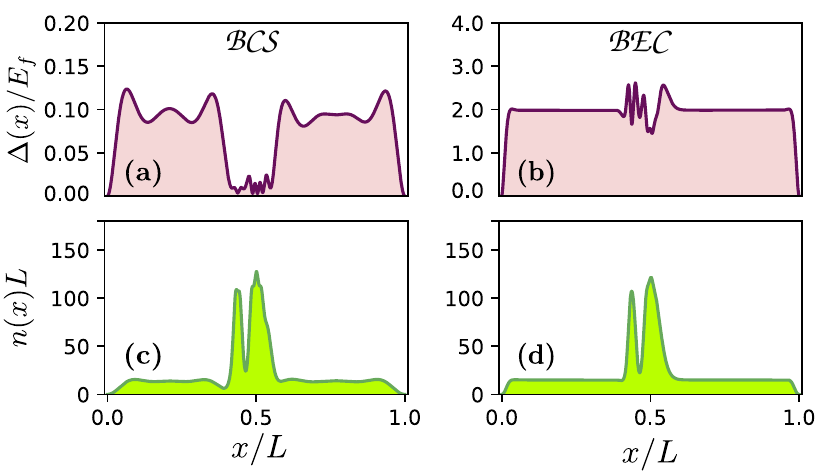}
\caption{Gap function $\Delta $ (top panels) and density (bottom panels) of a Fermi superfluid in the BCS regime with $g=0.7$, $\mu=0.20E_f$ (left panels) and the BEC regime where $g=3.6$ and $\mu=-0.46E_f$ (right panels) under the same Rydberg potential. Reduced (enhanced) undulations of $\Delta(x)$ near the wells of the Rydberg potential are evident in the BCS (BEC) regime.}
\label{Fig:Rbn62_Profile}
\end{figure}

\paragraph*{\textit{ Spectroscopic signatures of pair breaking and pair trapping}.--} 
To account for the impenetrable core of the Rydberg atom, the system is embedded in a 1D box of size $L$ with the Rydberg atom at $x=0$ and appropriately adjusting the relevant energy and length scales, see also SM~\cite{supp}. 
The gap function and density of a representative BCS (BEC) Fermi superfluid with $\mu>0$ ($\mu<0$) subjected to the Rydberg potential are depicted in the left (right) panels of Fig.~\ref{Fig:Rbn62_Profile}. 
While the density profiles on both BCS and BEC sides show peaks evidencing the bound states due to the attractive Rydberg potential, the most prominent contrast is the enhancement (suppression) of the gap function around the minima of the Rydberg potential in the BEC (BCS) regime. 
The decoupling of the gap function and density of a Fermi superfluid on the BCS side has also been discussed in vortex structures~\cite{PhysRevA.74.021602,PhysRevA.108.053303}.
The oscillatory boundary effects on the BCS side due to fermionic excitations are explained in the SM~\cite{supp}.

\begin{figure}[t]
\centering
\includegraphics[width=\columnwidth]{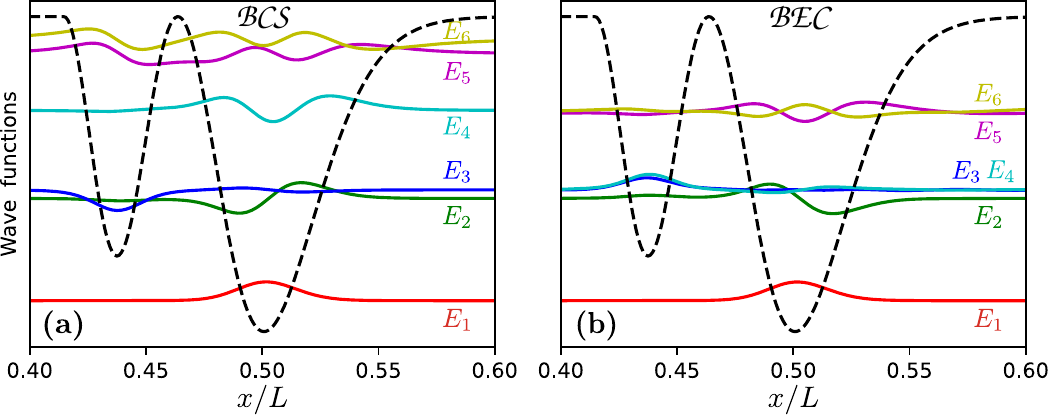}
\caption{Bound vibrational wave functions $v_n(x)$ of the Rydberg potential (dashed lines) depicted in Fig.~\ref{Fig:Demo} offset according to their energies $E_n$. 
The Fermi superfluid is in the (a) BCS regime with $g=0.7$ and $\mu=0.20E_f$ and (b) BEC regime with $g=3.6$ and $\mu=-0.46E_f$.
On the BEC side, there are two sets of nearly degenerate vibrational states localized respectively in the inner and outer wells, heralding the formation of triatomic heteronuclear Rydberg molecules.}
\label{Fig:Rbn62_BS}
\end{figure}

The bound-state wave functions $v_n(x)$ of the Rydberg potential in the BCS and BEC regimes are presented in Fig.~\ref{Fig:Rbn62_BS}, see SM~\cite{supp} for all bound-state wave functions $u_n$ and $v_n$.
Each well may host a series of bound states when the depth of the Rydberg potential is enough to compete with the pairing in the Fermi superfluid. 
Thus, there is a competition between the intercomponent fermion  attraction 
to maintain the Cooper pairs and the attraction among the Rydberg atom and the fermions to form molecules. 
The bound-state energies in the BCS regime are clearly  separated, and each bound state consists of a single fermion. This implies that the resulting diatomic Rydberg molecules originate from individual fermions due to broken Cooper pairs. 

The bound states in the BEC regime shown in Fig.~\ref{Fig:Rbn62_BS}(b) are more complex. Indeed, focusing on the furthest well, the first two bound states are clearly separated in energy, indicating that they correspond to diatomic Rydberg molecules. However, the subsequent two higher vibrational bound states in the same well are energetically adjacent with almost identical wave functions. Together with the enhanced gap function shown in Fig.~\ref{Fig:Rbn62_Profile}, the twin bound states suggest the presence of a locally trapped Cooper pair. Therefore, the furthest well hosts a triatomic Rydberg molecule as an excited vibrational state in the BEC regime due to the combination of the strong Cooper pairing and the Rydberg potential being capable of trapping the Cooper pair.  
There is also a pair of bound states with almost identical binding energies and wave functions localized in the secondary well illustrated in Fig.~\ref{Fig:Rbn62_BS}(b). 
These are again evident of the creation of another triatomic Rydberg molecule.
Therefore, the double-well approximate Rydberg potential depicted in Fig.~\ref{Fig:Demo} is able to host both diatomic and triatomic Rydberg molecules. 
Although the excited vibrational-state wave functions may extend into the inner potential wells, a four-well calculation, described in SM~\cite{supp}, confirms that the results  with two outermost wells are valid.

The Cooper-pair size may be estimated by the BCS coherence length~\cite{Leggett}
\begin{eqnarray}\label{Eq:xi}
\xi\approx \frac{\hbar v_f}{\Delta},
\end{eqnarray}
where $v_f$ is the Fermi velocity. For the system studied here, the full width at half maximum of the furthest (secondary) well is about $0.04L$ ($0.02L$). The Cooper-pair size of the selected BCS (BEC) case of Fig.~\ref{Fig:Rbn62_Profile} is $\xi/L\approx 0.06$ ($\xi/L\approx 0.003$) since $\Delta/E_f\approx0.10$ and $k_f L\approx 35$ ($\Delta/E_f\approx2.0$ and $k_f L\approx 36$). Hence, the Cooper pairs on the BCS side cannot be accommodated within the Rydberg-potential wells. In this context, only a fermion from a broken Cooper pair is captured, forming a diatomic molecule. In contrast, the Cooper pairs of the BEC case may fit into the Rydberg potential, which is deep enough to either break a Cooper pair or trap it to form a diatomic or a triatomic Rydberg molecule.

For a typical cold-atom cloud with density $n\approx 10^{14}$ cm$^{-3}$~\cite{Pethick-BEC}, $E_f\approx 10$kHz for Rb atoms, the depth of the Rydberg potential in Fig.~\ref{Fig:Demo} reaches the order of MHz. The pairing gap is roughly of the order of $E_f$ as shown in Fig.~\ref{Fig:Rbn62_Profile}, which can be orders of magnitude smaller than the depth of the Rydberg potential even on the BEC side. 
The Rydberg molecule lifetime is typically about 10-100 $\mu$s~\cite{Shaffer2018}, while the timescale in a Fermi gas is governed by $\hbar/E_f$ ($\sim$ 0.1ms). Therefore, the above treatment of quasi-equilibrium of a Fermi superfluid in the presence of Rydberg molecules is physically valid.
The Rydberg potential is seen as a spatially localized impurity to the Fermi superfluid, imprinting the resulting local deformation, before the global collective effects of the superfluid set in.
Moreover, since there are only few Rydberg atoms in a Fermi superfluid and the Rydberg potentials are local with finite lifetime, the feedback from the Rydberg-molecule formation on the Fermi superfluid, such as heating, is assumed to be negligible.
Meanwhile, a shallow Rydberg potential discussed in the SM~\cite{supp} is shown to also form diatomic and triatomic Rydberg molecules.

\begin{figure}
\centering
\includegraphics[width=\columnwidth]{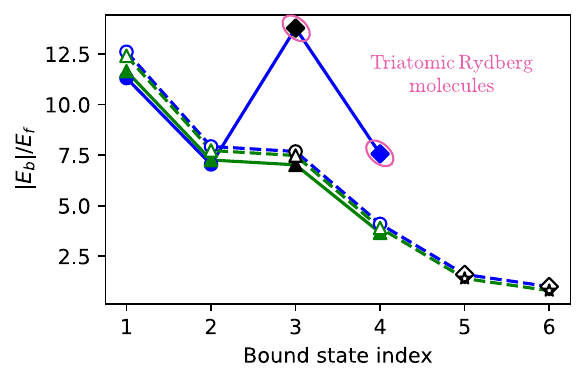}
\caption{Normalized binding energies from the BdG equation (circles and diamonds) and the Schr\"odinger equation (triangles) with the Rydberg potential of  Fig.~\ref{Fig:Demo}. Solid and hollow circles and triangles (diamonds) denote diatomic (triatomic) Rydberg molecules. Dashed (solid) lines connect the data on the BCS (BEC) side with $g=0.7$ and $\mu=0.20E_f$ ($g=3.6$ and $\mu=-0.46E_f$). Blue and green (black) symbols denote the bound states in the outer (inner) well of the Rydberg potential. The higher bound states with indices $5,6$ occupy both wells. Each triatomic Rydberg molecule binds a Cooper pair.}
\label{Fig:Eb_Rbn62}
\end{figure}

The respective binding energies (normalized by $E_f$) obtained from the BdG equation with the Rydberg potential of Fig.~\ref{Fig:Demo} are illustrated in Fig.~\ref{Fig:Eb_Rbn62}. 
The first two lowest-energy bound states in both BCS and BEC regimes have comparable binding energies since they correspond to diatomic Rydberg molecules consisting of the Rydberg atom and a broken-pair fermion. However, the binding energies of the higher vibrational bound states in the BCS and BEC regimes deviate more significantly because the triatomic Rydberg molecules in the BEC possess relatively larger binding energies within the trapped Cooper pairs. 

We remark that the relation between the diatomic and triatomic Rydberg molecules in Fermi superfluids is more complex than that in a BEC~\cite{PhysRevLett.120.083401} due to spin statistics and many-body effects. 
Indeed, adding an identical boson to a Rydberg dimer leads to a triatomic molecule with twice the diatomic binding energy. 
However, this does not hold for fermions due to the Pauli exclusion. 
Specifically, the formation of diatomic and triatomic Rydberg molecules in a Fermi superfluid competes with the binding of Cooper pairs. 
As such, the many-body contribution of breaking or trapping a Cooper pair plays a decisive role in creating Rydberg molecules, 
as it becomes apparent by the BdG calculation shown in Fig.~\ref{Fig:Eb_Rbn62}.

To discern many-body from single-particle  effects in the Rydberg molecule formation, we also evaluate the binding energies of diatomic Rydberg molecules with the Schr\"odinger equation $(h_\sigma+V_{Ryd})\psi_\sigma=E^{S}_n\psi_\sigma$, with $h_\sigma$ from $\mathcal{H}_{BCS}$; see also SM~\cite{supp} for the underlying bound states.
The normalized single-particle binding energies  are presented in Fig.~\ref{Fig:Eb_Rbn62}. 
The many-body binding energies obtained from the BdG equation are in general slightly larger than the corresponding single-particle energies due to pairing effect. However, the binding energies in the BCS regime follow a similar trend with the single-particle energies, and their energy difference remains roughly constant as higher vibrational states are reached. 
In contrast, the BdG binding energies on the BEC side exhibit larger deviations from their single-particle counterparts. 
The emergence of the triatomic Rydberg molecules results in a substantial energy difference from their noninteracting counterpart due to the trapped Cooper pair which keeps its own binding energy.

\paragraph*{\textit{Implications for experimental realization}.--} Spatially resolved rf spectroscopy of atomic Fermi superfluids~\cite{PhysRevLett.99.090403,doi:10.1126/science.aan5950}, following original attempts in Refs.~\cite{Kinnuen04,PhysRevLett.101.140403}, 
maps out the local pairing gap. As described in Fig.~\ref{Fig:Rbn62_Profile}, this will determine the types of Rydberg molecules since the pairing is suppressed (enhanced) in the diatomic (triatomic) Rydberg molecule. Meanwhile, the Rydberg molecules in a Fermi superfluid may serve as a probe for the Cooper-pair size because triatomic Rydberg-molecule formation is only possible when the Cooper-pair size is smaller than the width of the Rydberg potential. Differentiating the diatomic and triatomic Rydberg molecules is also achievable by Rydberg-molecule line spectroscopy~\cite{PhysRevLett.120.083401,PhysRevA.97.022707}. For example, the binding energies in the BCS (BEC) regime shown in Fig.~\ref{Fig:Eb_Rbn62} are $12.5,7.9,7.6,4,1,1.6,1.0$ MHz ($12,7.7,15,8.4$ MHz).  
At those values, red detuned spectroscopy of the Rydberg lines will show peaks, corresponding to the formation of oligomeric Rydberg molecules, see for instance Fig. 2 in Ref.~\cite{PhysRevLett.120.083401}.
The Rydberg impurity-Fermi superfluid system features several tunable parameters, including the depth, width, and location of the Rydberg potential, determined by the Rydberg excitation~\cite{Shaffer2018,Fey20}, and the pairing strength 
and particle density 
of the Fermi superfluid (see, e.g., Refs.~\cite{Pethick-BEC,Leggett}).

Furthermore, the quasi-1D setup has several advantages. First, the many-body lifetime induced by Rydberg atoms in a lattice is found to be longer for reduced dimensions~\cite{PhysRevX.7.041063}. If similar enhancement also holds in the continuum, 
it may facilitate Rydberg-molecule formation in 1D as there are on average few fermions within the Rydberg orbit, in the cases studied here. 
Second, the rotational excitations of Rydberg molecules will be less relevant in 1D, significantly simplifying the bound-state spectrum. Moreover, the 1D geometry eases i) the comparison between the Cooper-pair size and the Rydberg potential width, ii) the identification of diatomic or triatomic Rydberg molecules, and iii) the characterization of the Rydberg molecules, e.g., from the density and pairing gap profiles.

So far, the Rydberg atoms are assumed to be of different isotopes or species from the Fermi superfluid. We envision that future experiments similar to Refs.~\cite{PhysRevA.62.011402,Onofrio16} will prepare a boson-fermion mixture, excite the bosons to Rydberg states, produce Rydberg molecules in the Fermi superfluid, and measure the pairing gap and binding energy by spatially resolved rf spectroscopy and molecular line spectroscopy, respectively. Alternatively, if some of the fermions within the superfluid are excited into Rydberg atoms, forming homonuclear Rydberg molecules, this results in a reduced effective pairing gap, see the SM~\cite{supp}.
Once the excited Rydberg atoms are present, however, the corresponding bound states can be extracted through the BdG formalism.  Therefore, dimer or trimer Rydberg molecules are expected via Rydberg excitations stemming from the Fermi superfluid although the reduced effective pairing gap will favor dimer Rydberg molecules.

Finally, we note that Rydberg molecules are different from Cooper-pair splitting in superconductor heterostructures~\cite{Lesovik2001,PhysRevB.63.165314,Hofstetter2009,Ranni2021,PhysRevB.109.L081402,PhysRevLett.109.157002} In this case, 
the proximity effect is utilized by dynamically sending a Cooper pair, as an excited state with spin entanglement or momentum correlation, to two separate non-superconducting regions in real space.
In contrast, the fermion bound in a diatomic Rydberg molecule no longer retains the pairing correlation, while the tightly-bound Cooper pair in a triatomic Rydberg molecule localizes in real space. 
Along the same lines, there are subtle differences between the Rydberg molecules in Fermi superfluids and the binary tidal disruption event and pion matter. For instance, binding in binary stars (pion matter) stems from gravity (Coulomb interactions), whereas in Rydberg molecules, it is traced back to the electron-atom scattering.

\paragraph*{\textit{Summary and outlook}.--}The bound states of Fermi superfluids in a Rydberg-impurity potential testify the formation of Rydberg molecules. The tunable fermion pairing gives rise to diatomic (triatomic) Rydberg molecules from broken (tightly-bound) Cooper pairs, exhibiting different features of the gap function due to their distinctive nature. The detection of the triatomic Rydberg molecules may reveal information about the Cooper-pair size, while the bound-state energies reflect pairing effects.
With the rapid developments of Rydberg physics and Fermi gases, realizations of Rydberg molecules in Fermi superfluids will provide an elegant example of interfacing few- and many-body physics.
Furthermore, going beyond the Leggett-BCS theory~\cite{Pethick-BEC,Leggett} of the superfluid ground state, pre-formed Cooper pairs at finite temperatures correct the superfluid transition temperature and lead to the pseudo-gap effect away from the BCS regime~\cite{Gaebler2010,Li2024}. Incorporating pairing-fluctuation theories developed for homogeneous systems~\cite{LEVIN2010233,ZwergerBook,RanderiaReview,Mueller_2017} into the BdG formalism remains a challenge, and finite-temperature physics of Rydberg molecules awaits future research.

\paragraph*{\textit{ Acknowledgements}.--}C. C. C. was partly supported by the NSF (No. PHY-2310656). The authors appreciate the calculation of Rb Rydberg wave functions by Mariusz Pawlak. Support for ITAMP by the NSF is acknowledged. 

%


\onecolumngrid
\newpage
\clearpage
\begin{center}
\large{\bf{Supplemental Material: Breaking and trapping Cooper pairs by Rydberg-molecule spectroscopy in atomic Fermi superfluids}}
\end{center}
\twocolumngrid
\setcounter{equation}{0}
\setcounter{figure}{0}
\setcounter{section}{0}
\setcounter{page}{1}
\makeatletter
\renewcommand{\theequation}{S\arabic{equation}}
\renewcommand{\thefigure}{S\arabic{figure}}



\section{Bogoliubov-de Gennes calculation} 
After diagonalization by the BdG transformation (see the main text), the many-body Hamiltonian takes the form 
$\mathcal{H}={\sum_{\tilde{n}w}}' E_{\tilde{n}w} \gamma_{\tilde{n}w}^\dagger \gamma_{\tilde{n}w} +E_g$.
In this expression, $w=1,2$ represents the quasi-particle components, $E_g$ is the ground-state energy, and ${\sum_{\tilde{n}w}}'$  denotes summation over the positive-energy states.
The ground-state energy is $E_g=-\frac{|\Delta|^2}{U}+\sum_{\tilde n,w}(\epsilon_{\tilde nw}-E_{\tilde{n}w})$ with
$\epsilon_{\tilde {n}w}$ being 
the non-interacting counterpart of the excitation energy $E_{\tilde {n}w}$. 
The BdG equation has the symmetry
$\begin{pmatrix}
    u^{\tilde{n}2}_{\downarrow}(x)\\
    v^{\tilde{n}2}_{\uparrow}(x)
    \end{pmatrix}
    =\begin{pmatrix}
    v^{\tilde{n}1 *}_{\downarrow}(x)\\
    -u^{\tilde{n}1}_{\uparrow}(x)
    \end{pmatrix}$
with $E_{\tilde{n}2}=-E_{\tilde{n}1}$.
The quasi-particle operators obey $\langle\gamma_{\tilde{n}w}^\dagger\gamma_{\tilde{m}v}\rangle=\delta_{\tilde{n}\tilde{m}}\delta_{wv}f(E_{\tilde{n}w})$ and $ \langle\gamma_{\tilde{n}w}\gamma_{\tilde{m}v}\rangle=\langle\gamma_{\tilde{n}w}^\dagger\gamma_{\tilde{m}v}^\dagger\rangle=0$ with $f(E)=[e^{E/k_B T}+1]^{-1}$
being the Fermi distribution function. At finite temperatures, the gap function becomes
$\Delta(x)=-U{\sum_{\tilde{n}}}' u^{\tilde{n}}_{\uparrow }(x)v^{\tilde{n}*}_{\downarrow}(x)\tanh(E_{\tilde{n}}/k_B T)$ and the total Fermi density $n(x)=\sum_\sigma n_{\sigma}(x)=2{\sum_{\tilde n }}' [|v_{\tilde n}(x)|^2(1-f(E_{\tilde{n}}))+|u_{\tilde n}(x)|^2 f(E_{\tilde{n}})]$. We caution that a bound state with $E_n < 0$ contributes to the density via $|u_n|^2$. Due to the symmetry, it is equivalent to a $E_n > 0$ state contributing to the density via $|v_n|^2$.

In our numerical calculations, we consider a quasi-1D system in a 1D box of length $L$. We discretize the space $x/L=[0,1]$ using $n_x$ grid points $x_j=j \delta x$, where $\delta x=L/n_x$ and $j=0,1,2,....,n_x-1$. The BdG equation is also discretized by using the finite-difference method and becomes
\begin{equation}\label{Eq:DiscreteBdG}
   \sum_j \begin{pmatrix}
        h_{ij}&\Delta_{ij}\\
        \Delta^*_{ij}& -h_{ij}
    \end{pmatrix}
    \begin{pmatrix}
        u_j^{\tilde{n}}\\
        v_j^{\tilde{n}}
    \end{pmatrix}
    =E_{\tilde{n}}\begin{pmatrix}
        u_i^{\tilde{n}}\\
        v_i^{\tilde{n}}
    \end{pmatrix}.
\end{equation}
Here $\Delta_{ij}=\Delta_i\delta_{i,j}$ for $s$-wave pairing. The BdG Hamiltonian has the size of $2n_x\times 2n_x$ and we only take the positive energy eigenstates for the calculations of the gap function and density. For the ground state of the Fermi superfluid, the total density 
becomes
\begin{equation}\label{Eq:rhoBdG}
n(x)=2{\sum_{\tilde n }}' |v_{\tilde n}(x)|^2.
\end{equation}
The total fermion number is $N=N_{\uparrow}+N_{\downarrow}=\int_0^L n(x)dx$.
The gap function is given by
\begin{equation}\label{Eq:DeltaBdG}
    \Delta(x)
    =-U{\sum_{\tilde n}}' u_{\tilde n}(x)v_{\tilde n}(x).
\end{equation}

The Rydberg atom is placed at $x=0$, where the wave functions should vanish
due to the impenetrable core of the Rydberg atom. 
The box boundary at $x=L$ is chosen such that the wave functions return to the bulk values before encountering the wall at $x=L$.
The box introduces an energy scale $E_0=\hbar^2/(2mL^2)$, but we use the intrinsic inverse-length and energy units $k_f$ and $E_f$ carried by the fermions. 
Numerically, we start with a trial $\Delta(r)$ and a given set of parameters $(U,\mu)$ in order to solve the BdG equation and thus  obtain the eigenvalues and eigenstates of the Rydberg atom-Fermi superfluid system. The gap function is then assembled for the next iteration. The iteration stops when the convergence condition $(1/L)\int_0^L dx ||\Delta^{new}(x)|-|\Delta^{old}(x)||/E_0 < \epsilon$ is satisfied, where $\Delta^{new/old}(x)$ denote the gap functions between consecutive iterations. We have taken $\epsilon=10^{-6}$ and $1000$ grid points and checked that further adjustments of those values do not cause qualitative changes.

Let us also comment on the oscillatory boundary effects due to the box confinement on the BCS side in the profiles shown in Fig.~\ref{Fig:Rbn62_Profile} of the main text. The Fermi superfluid on the BCS side has relatively weak pairing. Therefore, when the boundary enforces the gap function and density to vanish, the fermions behave as  noninteracting ones exhibiting an  oscillatory behavior with length scale $1/k_f$. This is because noninteracting fermions form a Fermi sea and the perturbation of the system starts at the Fermi momentum, which then results in the aforementioned oscillations at the boundary. In contrast, the fermions form tightly bound pairs on the BEC side, which behave like composite bosons and no longer follow the Fermi statistics. On the BEC side, the composite bosons are repelled by the box boundaries, but they vanish smoothly without the oscillatory behavior. We emphasize that the box confinement is a simple choice to satisfy the impenetrability condition of the Rydberg atom, and the focus should be on the Rydberg-molecule formation due to the Rydberg potential instead of the boundary effects due to the choice of the confinement.

\begin{figure}
\centering
\includegraphics[width=\columnwidth]{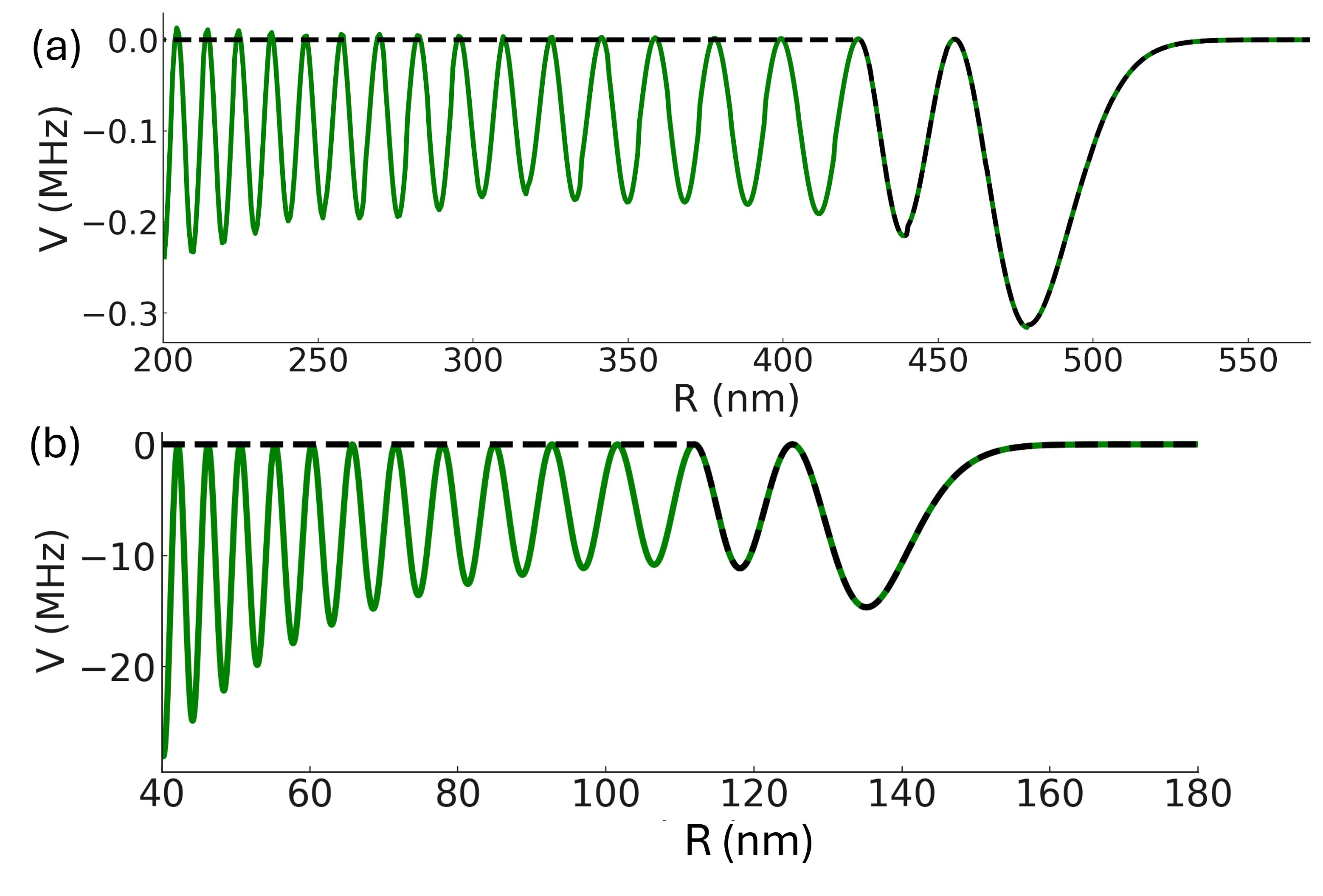}
\caption{Typical Rydberg potentials from (a) Sr(71S) state and (b) Rb(40S) Rydberg excitation. The solid (dashed) lines are the full potentials (outer double-well approximations). }
\label{Fig:Rbn62pot}
\end{figure}

\section{Higher Rydberg excitation}
Upon considering a higher Rydberg excitation results in a shallower Rydberg potential. As a paradigmatic example, here, we assume the Sr(71S) Rydberg state in a Fermi superfluid. Since the most relevant part of the Rydberg potential corresponds to the two dips furthest away from its core, we smooth out the highly oscillatory potential located close to the core. A comparison of the Rydberg potentials from the Rb(40S) state and that from the Sr(71S) state is given in Fig.~\ref{Fig:Rbn62pot}, along with the approximation concerning their two furthest potential wells used in the BdG calculations. When the two potentials are placed in a 1D box with the furthest dips at its center and the length and energy scales properly scaled, as explained below, the potential from the Rb(40S) state is about ten times deeper. The two Rydberg potentials thus allow us to contrast the characteristics of the emergent Rydberg molecules in Fermi superfluids.

Placing the approximate double-well Rydberg potential in a 1D box accounts for the impenetrable core of the Rydberg atom and provides a consistent comparison when different Rydberg states are used. We extract the distance $R_0$ from the core to the furthest dip and its depth $V_0$ of the Rydberg potential. For a selected Rydberg atom and its state, $R_0$ and $V_0$ are related. For example, $R_0=480$ nm and $V_0=0.32$ MHz for the Sr(71S) Rydberg state in a $^{87}$Sr Fermi superfluid. 
The aspect ratio of the Rydberg potential is fixed by introducing the energy scale $E_R=\hbar^2/(2mR_0^2)$ and obtaining the ratio $V_0/E_R$. By setting $R_0=\alpha L$ with $0<\alpha < 1$, the furthest dip is located at $\alpha L$ inside the box. This also fixes the energy relation $E_R=E_0/\alpha^2$, from which the Rydberg potential can be expressed in terms of the corresponding $k_f$ and $E_f$.

\begin{figure}[t]
\includegraphics[width=\columnwidth]{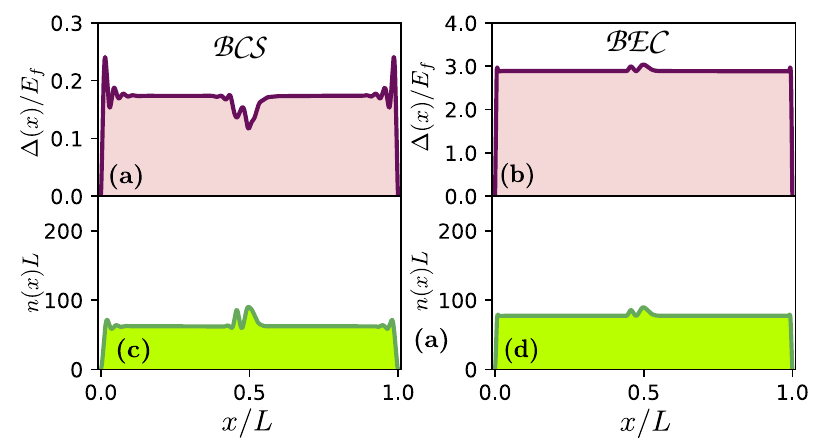}
\caption{Profiles of the gap function (top row) and density (bottom row) of a Fermi superfluid in the BCS regime with $g=1.0$ and $\mu=0.52E_f$ (left column) and BEC regime with $g=4.1$ and $\mu=-0.07E_f$ (right column). 
The Rydberg potential stems from the Sr(71S) Rydberg state and its furthest dip is located at $R_0=L/2$.}
\label{Fig:Srn71_Profiles}
\end{figure}

By squeezing the Rydberg potential towards the left of the box with a smaller $\alpha$, the effective depth of the potential increases, but the widths of the wells decrease. As shown below, the case with $\alpha=1/4$ exhibits similar behavior on the BCS and BEC sides as those where $\alpha=1/2$ being presented here. This indicates that the number of bound states is determined by a combination of the potential width and depth. Thus, a simple rescaling of the Rydberg potential inside the box does not lead to further qualitative changes. In the following, we will place the furthest dip of the Rydberg potential at $x=L/2$ and scale its depth accordingly.

The left column of Figure~\ref{Fig:Srn71_Profiles} shows the profiles of the gap function and the density for a selected case in the BCS regime  with $\mu>0$ and the potential taken from the Sr(71S) Rydberg state. The presence of the Rydberg potential causes a dip in the gap function, which implies a suppression of pairing, and a peak in the density, indicating that unpaired fermions accumulate.

\begin{figure}[t]
\includegraphics[width=\columnwidth]{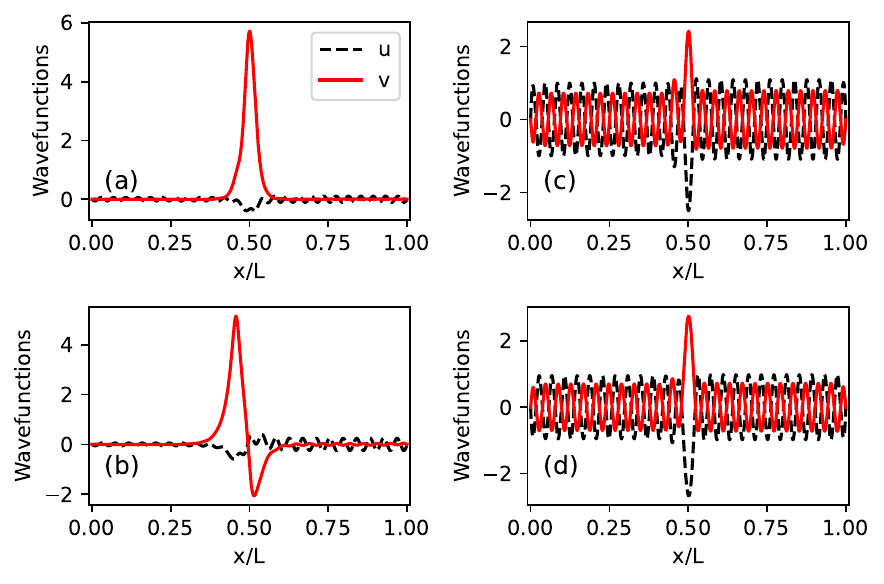}
\caption{Bound-state wave functions of a Fermi superfluid under the Sr(71S) Rydberg potential in the BCS regime with $g=1.0$ and $\mu=0.52E_f$ (left panels) and BEC regime with $g=4.1$ and $\mu=-0.07E_f$ (right panels). The bound states on the BCS side are of broken-pair nature while the two on the BEC side are from a trapped Cooper pair. 
The furthest well of the Rydberg potential is at $R_0=L/2$.}
\label{Fig:Ryd05_BS}
\end{figure}

Before analyzing the energy spectrum and wave functions of the BCS case, we show in the right column of Fig.~\ref{Fig:Srn71_Profiles} the profiles of the gap function and density of a Fermi superfluid on the BEC side with $\mu<0$ under the same Rydberg potential. In stark contrast to the BCS case, the gap function features a peak at the location of the Rydberg potential. The same behavior is evident on the 
density profile. Thus, the pairing in enhanced inside the Rydberg potential when the Fermi superfluid is BEC-like. The enhancement of pairing also suggests that Cooper pairs are trapped by the Rydberg atom.

To explain the contrast between the BCS and BEC cases, we analyze the eigen-functions obtained from the BdG equation, see  Fig.~\ref{Fig:Ryd05_BS}. Since the symmetry of the BdG equation guarantees that every positive energy eigen-state is accompanied by a negative-energy state, a bound state of a particle is also accompanied by a bound state of a hole having the opposite energy sign. 
For this reason, we examine the profiles of the eigen-functions and identify the bound states with localized patterns in the potential wells. 
For the BCS and BEC cases presented in Fig.~\ref{Fig:Srn71_Profiles}, most of the eigen-states exhibit oscillatory behavior in the box (not shown for brevity). Nevertheless, we identify particular states featuring localization at the dips of the Rydberg potential. The left column of Fig.~\ref{Fig:Ryd05_BS} depicts the two energetically lowest localized states on the BCS side that can be identified with this Rydberg potential. From the number of nodes inside the Rydberg potential, one may classify them as the first and second bound states localized in the rightmost potential well.

The binding energies of these two bound states on the BCS regime are not adjacent to each other, implying that they are between a fermion and the Rydberg atom as only individual fermionic states well separated in energy are involved. 
This behavior suggests that the composite object refers to a diatomic Rydberg molecule.
The suppression of the gap function on the BCS side can also be understood in terms of the formation of the Rydberg molecules, which breaks a Cooper pair when a fermion falls into the Rydberg potential. Meanwhile, the fermionic atom from a broken Cooper pair in the Rydberg molecule causes a bump in the local density, reminiscent to the occupation of the vortex core by unpaired fermions~\cite{PhysRevA.74.021602,PhysRevA.108.053303}.

For the BEC case demonstrated in the right column of Fig.~\ref{Fig:Srn71_Profiles}, there are also two bound states localized in the Rydberg potential. However, the two localized states have adjacent energies and similar wave functions, as can be seen in the right column of Fig.~\ref{Fig:Ryd05_BS}. Therefore, two fermions are constituting the first bound state in the Rydberg potential in the BEC regime. This is a direct indication that a tightly-bound Cooper pair falls into the Rydberg potential and forms a three-body bound state, or a Rydberg molecule with a Rydberg atom and a Cooper pair.  Therefore, a triatomic Rydberg molecule forms in the BEC case presented here.
For the Sr(71S) Rydberg potential, however, the depth is too shallow to host additional bound states.

We also estimate the Cooper-pair size using Eq.~\eqref{Eq:xi} in the main text and compare it with the width of the Rydberg potential.
For the Sr(71S) Rydberg state, the width of the furthest well is about $0.04L$ after placing it at $x/L=0.5$ of the box. For the BCS case selected above, $\Delta/E_f\approx 0.18$ and $k_f L\approx 98$, which lead to $\xi/L\approx 0.06$. Meanwhile, $\Delta/E_f\approx 2.9$ and $k_f L\approx 121$ for the shallow-BEC case selected above, so $\xi/L\approx 0.003$. Therefore, the width of the Cooper pairs on the BCS (BEC) side is larger (smaller) than the width of the primary Rydberg-potential well. This also corroborates that a fermion from a broken Cooper pair is trapped by the Rydberg potential to form a diatomic Rydberg molecule when the Fermi superfluid is on the BCS side. In contrast, the relatively shallow Rydberg potential traps a Cooper pair on the BEC side in its primary well, forming  ``a Cooper pair in a molecule". 
Moreover, as compared to the relatively deep Rydberg-potential discussed  
in the main text, the secondary well of the 
shallower Rydberg potential presented here does not support bound states.

\begin{figure}[t]
\centering
\includegraphics[width=\columnwidth]{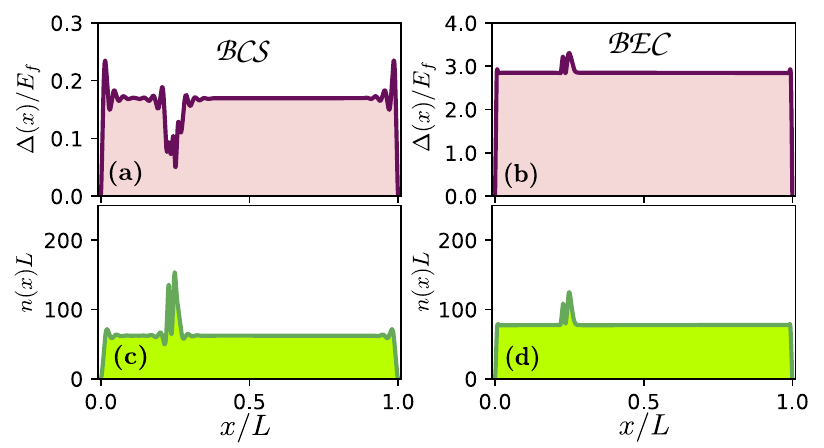}
\caption{Profiles of the  gap function (top row) and density (bottom row) of a Fermi superfluid in the BCS regime with $g=1.0$ and $\mu=0.52E_f$ (left column) and the BEC regime with $g=4.1$ and $\mu=-0.07E_f$ (right column). The potential is created by the Sr(71S) Rydberg state and has its furthest dip at $R_0=L/4$.}
\label{Fig:Ryd025Profile}
\end{figure}

\section{Adjusting the location of the Rydberg potential}

Here we take the potential of the Sr(71S) Rydberg state but set $R_0=L/4$, i.e., $\alpha=1/4$. 
Using this scaling, the depth of the potential increases due to the relatively smaller $R_0$, but the width of the potential reduces.
Fig.~\ref{Fig:Ryd025Profile} presents the profiles of the  gap function and the density on the BCS side and BEC side, respectively, under the Sr(71S) Rydberg potential. Similar to the $\alpha=1/2$ case, the pairing is suppressed on the BCS side but is enhanced on the BEC side. Meanwhile, the density exhibits a peak on both BCS and BEC sides, signaling the emergence of bound states.

\begin{figure}[t]
\includegraphics[width=\columnwidth]{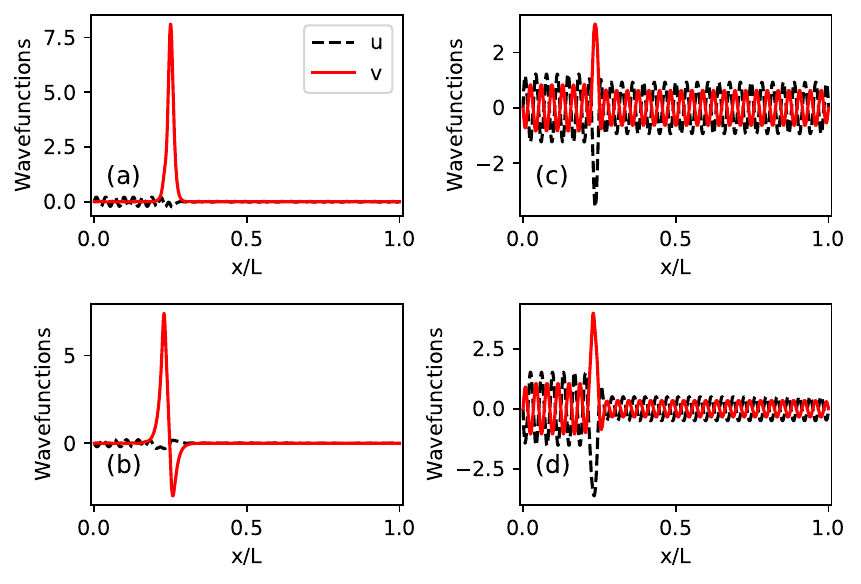}
\caption{Bound-state profiles of a Fermi superfluid upon formation of the Sr(71S) Rydberg potential in the BCS regime with $g=1.0$ and $\mu=0.52E_f$ (left column) and the BEC regime with $g=4.1$ and $\mu=-0.07E_f$ (right column). The outermost well of the Rydberg potential is located at $R_0=L/4$.}
\label{Fig:Ryd025_states}
\end{figure}

The left column of Fig.~\ref{Fig:Ryd025_states} illustrates  the localized eigenstates extracted from the BdG equation corresponding to the setup on the BCS side depicted in Fig.~\ref{Fig:Ryd025Profile}. It becomes apparent that there are again two bound states with clearly separated binding energies. By examining the nodes of the wave functions, we can infer that they are the first and second bound states of the Rydberg potential. Therefore, the Rydberg atom breaks a Cooper pair and captures a fermionic atom to form a diatomic Rydberg molecule.

In contrast, when the Fermi superfluid is in the BEC regime, the right column of Fig.~\ref{Fig:Ryd025_states} shows the localized eigenstates corresponding to the right column of Fig.~\ref{Fig:Ryd025Profile}. There are two bound states with adjacent binding energies and similar wave functions, indicating that a Cooper pair has been trapped by the Rydberg potential. Therefore, the triatomic Rydberg molecule in this case consists of the Rydberg atom and a Cooper pair trapped by its potential. Squeezing the Rydberg potential within the box increases its depth and reduces its 
width, but it is still not sufficient to break a tightly bound Cooper pair in the shallow BEC regime in this case.

\section{Details of bound states}
\begin{figure}[t]
\includegraphics[width=\columnwidth]{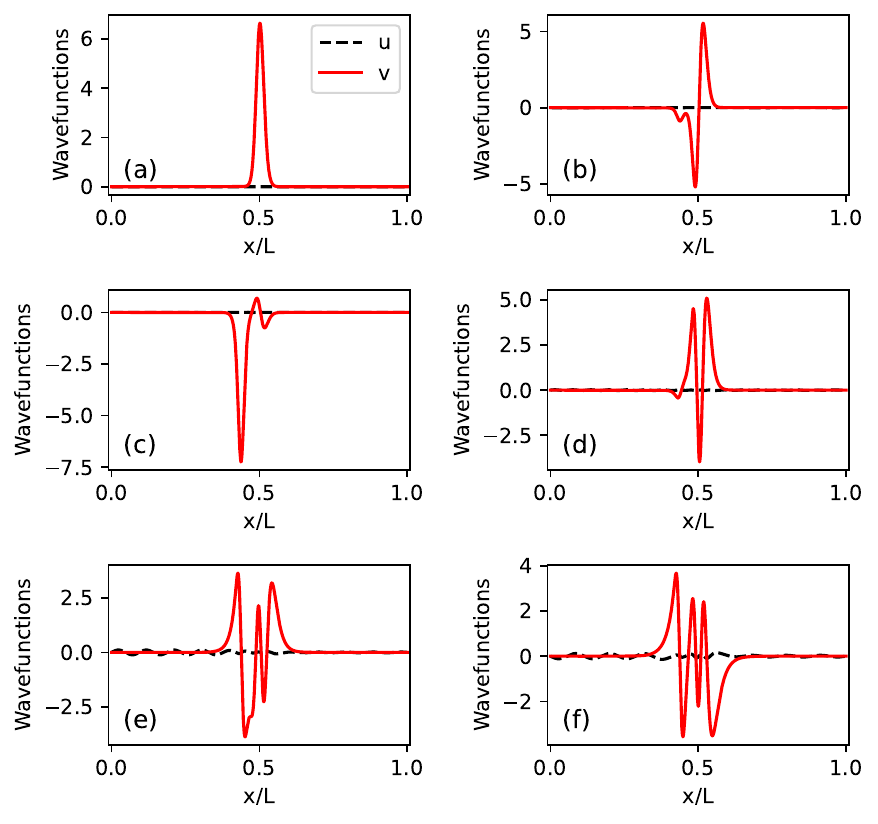}
\caption{Bound-state profiles of a Fermi superfluid in the BCS regime with $g=0.7$ and $\mu=0.20E_f$ with the Rb(40S) Rydberg potential. The magnitude of binding energies decreases monotonically from (a) to (f). Here panels (a), (b), and (d) show the wave functions for the three lowest bound states in the outermost well, while the wave function in panel (c) refers to the first bound state in in the inner well. Panels (e) and (f) are bound states occupying both wells. All states here stem from broken-pair fermions.}
\label{Fig:Rbn62BCS_states}
\end{figure}

The bound state wave functions $v_n(x)$ of the Rydberg potential are summarized in
Fig.~\ref{Fig:Rbn62_BS} of the main text. Here, for reasons of completeness, we provide the full bound-state wave functions $u_n$ and $v_n$ of the Fermi superfluid under the Rydberg potential shown in Fig.~\ref{Fig:Demo} of the main text.
Specifically, Fig.~\ref{Fig:Rbn62BCS_states} presents the localized eigenstates in the Rydberg potential which are extracted from the BdG analysis in the BCS case. Panels (a) to (f) depict, in a hierarchical order, the lowest-energy bound state to the highest-energy one. Since the bound-state energies are well separated and each bound state corresponds to a fermion, the bound states of Fig.~\ref{Fig:Rbn62BCS_states} are indicative of diatomic molecule formation.

\begin{figure}[t]
\includegraphics[width=\columnwidth]{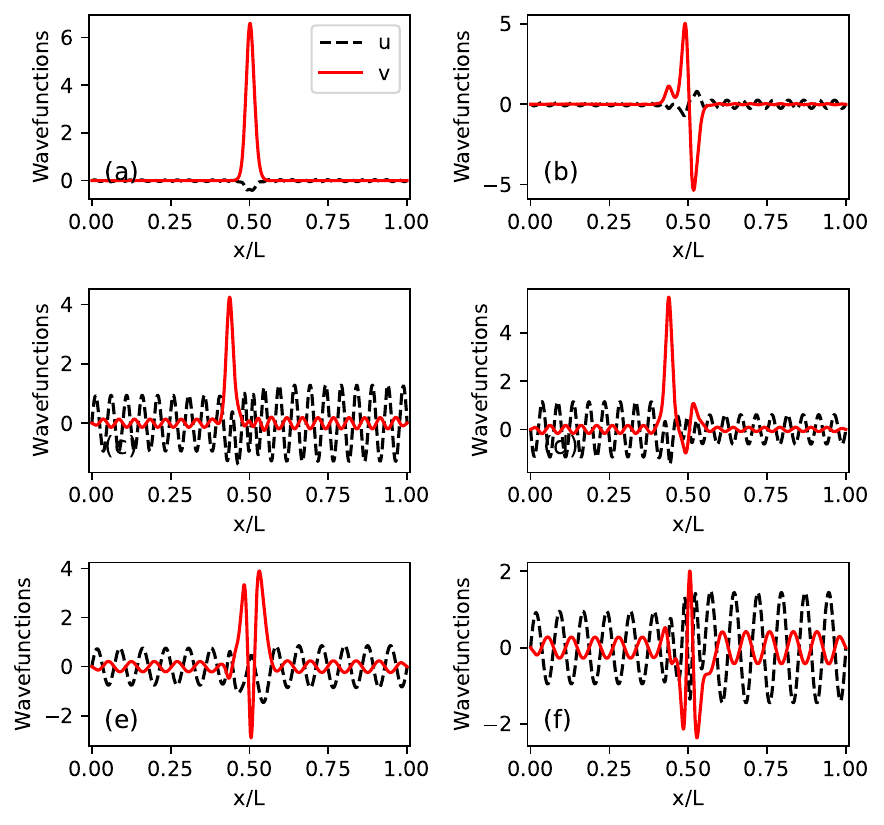}
\caption{Bound-state profiles of a Fermi superfluid in the BEC regime with $g=3.6$ and $\mu=-0.46E_f$ with the Rb(40S) Rydberg potential. The magnitude of binding energies decreases monotonically following panels (a) to (f). Here panels (a) and (b) depict the first two diatomic Rydberg-molecule states in the furthest well, panels (c) and (d) correspond to a triatomic molecule state in the inner well, and the states in panels (e) and (f) form another triatomic molecule state in the outermost well.
}
\label{Fig:Rbn62BEC_states}
\end{figure}

Fig.~\ref{Fig:Rbn62BEC_states} shows the 
bound states calculated from the BdG equation in the BEC regime utilizing the same Rydberg potential. 
As can be seen, there are two energetically distinct bound states localized in the furthest well [Fig.~\ref{Fig:Rbn62BEC_states}(a), (b)] followed by 
two almost identical lowest-energy bound states localized in the secondary well [Fig.~\ref{Fig:Rbn62BEC_states}(c), (d)] and another two almost identical bound states localized in the furthest well [Fig.~\ref{Fig:Rbn62BEC_states}(e), (f)]. Each pair of the twin bound states corresponds to a Cooper pair trapped by the Rydberg potential. However, the secondary well on the left traps a Cooper pair into its lowest bound state [Fig.~\ref{Fig:Rbn62BEC_states}(c), (d)], while the furthest well traps a Cooper pair as its third vibrational state [Fig.~\ref{Fig:Rbn62BEC_states}(e), (f)].

\section{Beyond double-well approximation}
In the main text, we assumed the double-well approximation for the Rydberg potential, see Fig. 1(a). This is expected to be a reasonable treatment since most of the excitation amplitude reside in the outer lobes of the Rydberg wave function which forms the outer wells. To confirm the validity of this approximation, here we consider the four furthest wells of the Rb(40S) potential shown in Fig. 1 (a), and perform the BdG calculations with the same parameters. Overall, as expected, additional higher energy bound states are identified in the third and fourth wells, which are typically of higher energies. The bound states of relevance in the furthest two wells are virtually the same as those obtained in the double-well approximation with their binding energies only slightly shifted.

\begin{figure}[t]
\includegraphics[width=\columnwidth]{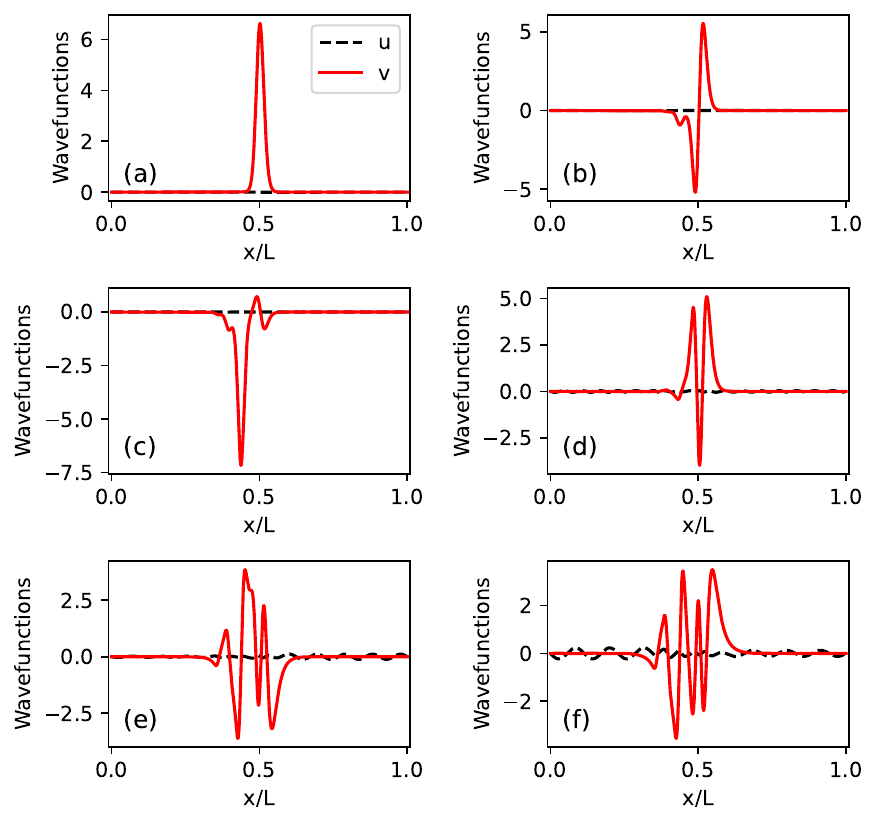}
\caption{Bound-state profiles of a Fermi superfluid in the BCS regime with $g=0.7$ and $\mu=0.23E_f$ using the four-well approximation of the Rb(40S) Rydberg potential. The magnitude of binding energies decreases monotonically following panels (a) to (f). Panels (a), (b), (d) refer to the lowest three bound states on the furthest well, panels (c) represents the lowest bound states on the second well, and panels (e) and (f) are bound states occupying the two furthest wells. The bound states in the two furthest wells agree with those from the double-well approximation shown in Fig.~\ref{Fig:Rbn62BCS_states}.}
\label{Fig:Rbn62FWBCS_states}
\end{figure}

\begin{figure}[t]
\includegraphics[width=\columnwidth]{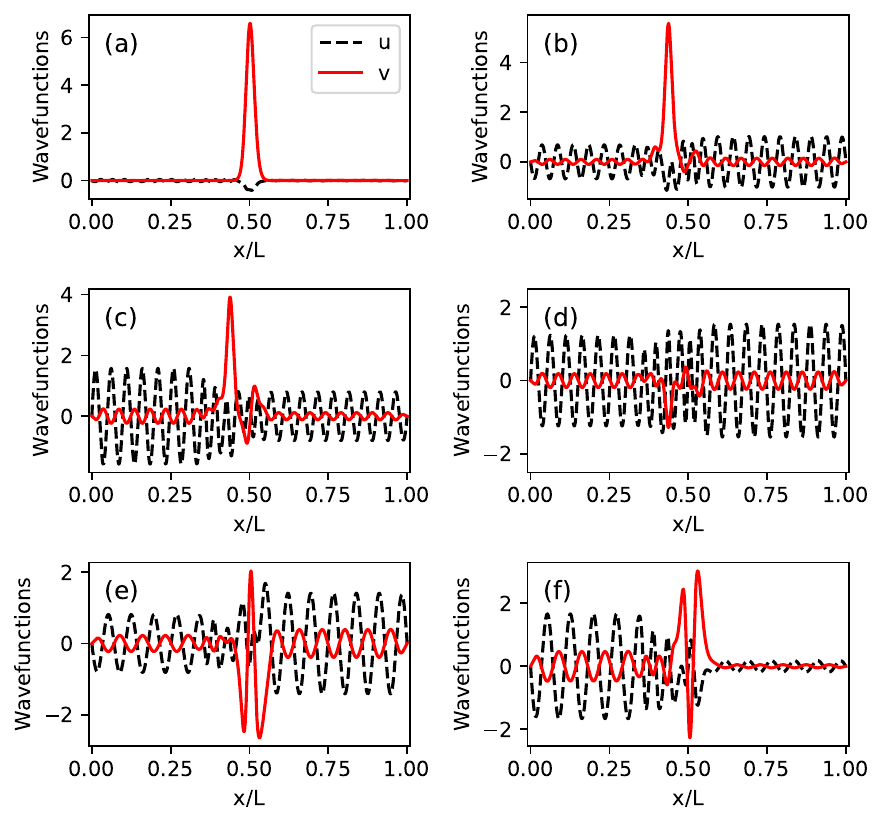}
\caption{Bound-state profiles of a Fermi superfluid in the BEC regime with $g=3.6$ and $\mu=-0.49E_f$ employing the four-well approximation for the Rb(40S) Rydberg potential. The magnitude of binding energies decreases monotonically following panels (a) to (h). Here, panels (a) and (d) are the two lowest diatomic states in the outermost well, panels (b) and (c) show the triatomic state in the second (first inner) well, panels (e) and (f) show the lowest triatomic state in the furthest well. The bound states in the outermost two wells agree with the results of the double-well approximation depicted in Fig.~\ref{Fig:Rbn62BEC_states} except a slight re-ordering. } 
\label{Fig:Rbn62FWBEC_states}
\end{figure}

To compare the double-well and four-well approximations, we extract the bound states within the four-well model in both the BCS and BEC regimes. Fig.~\ref{Fig:Rbn62FWBCS_states} presents the wave functions of the first six lowest bound states on the BCS side, which are almost identical to those from the double-well approximation shown in Fig.~\ref{Fig:Rbn62BCS_states}. The bound states in the third or fourth inner wells lie at higher energies. 
Next, we show the six lowest bound-state wave functions from the four-well approximation on the BEC side in Fig.~\ref{Fig:Rbn62FWBEC_states}. The low-energy bound states in the furthest two wells are virtually the same as those from the double-well approximation depicted in Fig.~\ref{Fig:Rbn62BEC_states}. A minor difference occurring is that the second diatomic bound state in the furthest well is slightly lower  than the first triatomic bound state in the second well in the double-well approximation. This is because those bound-states are already close in energy and their order is interchanged due to the quantitative changes in the presence of the additional inner wells. Again, the bound states in the inner wells beyond the two furthest wells have higher energies. Therefore, the double-well approximation of the Rydberg potential already captures all important features of the low-lying bound states.

\section{Exciting Rydberg atoms within the Fermi superfluid} 
Here, instead of introducing bosonic isotopes or atoms of different species as Rydberg impurities in an atomic Fermi superfluid, we consider the alternative process of exciting the atoms of the Fermi superfluid to produce Rydberg atoms. 
In this case, our description should be modified in order to account for the excitation of Rydberg atoms from the Fermi superfluid. This process is modelled by means of considering the additional Hamiltonian term $H_{ex}=g_l \int dx (d^\dagger \tilde{\psi}^\dagger(x)\psi_{\downarrow}(x)\psi_{\uparrow}(x) + h.c.)$ which, in practice, converts a Cooper pair into a Rydberg atom and an unpaired fermion. 
In this expression, $\tilde{\psi}^\dagger$ is the  creation operator of a fermion from the broken Cooper pair but without excitation to the Rydberg state, $d^{\dagger}$ refers to the creation operator of a Rydberg atom, and $g_l$ is the coupling constant for exciting the Rydberg atoms. Since a Cooper pair needs to be broken to create a Rydberg atom in this case, $g_l$ is expected to be larger than the pairing coupling constant $U$. 

If the Rydberg-atom density $n_R$ from the broken Cooper pairs satisfies $n_R R_0^3 \ll 1$, with $R_0$ denoting the range of the Rydberg potential, we may approximate $\langle d^\dagger \tilde{\psi}^\dagger\rangle \approx \langle d^\dagger d\rangle = n_R = \int dx\langle\tilde{\psi}^\dagger(x)\tilde{\psi}(x)\rangle$. Following this crude approximation, the excitation Hamiltonian becomes $H_{ex}\approx g_{l}n_R\int dx(\psi_{\downarrow}(x)\psi_{\uparrow}(x)+h.c.)$. This has a form similar to the pairing terms in the BCS Hamiltonian. In particular, if it is combined with the BCS Hamiltonian, the resulting Hamiltonian leads to a suppressed pairing gap $\Delta' = \Delta - g_l n_R$, where $\Delta$ is the pairing gap in the absence of the Rydberg excitation. Therefore, exciting the Rydberg atoms directly from the Fermi superfluid instead of introducing different isotopes or species results in a suppression of the pairing gap and favors the formation of a diatomic Rydberg molecule with a fermion from a broken Cooper pair. Moreover, the Rydberg molecules are homonuclear if the Rydberg atoms are excitations of the Fermi superfluid. In contrast, the Rydberg molecules discussed in the main text are heteronuclear because the Rydberg atoms are different from the fermions in the superfluid.

\begin{figure}[t]
\centering
\includegraphics[width=\columnwidth]{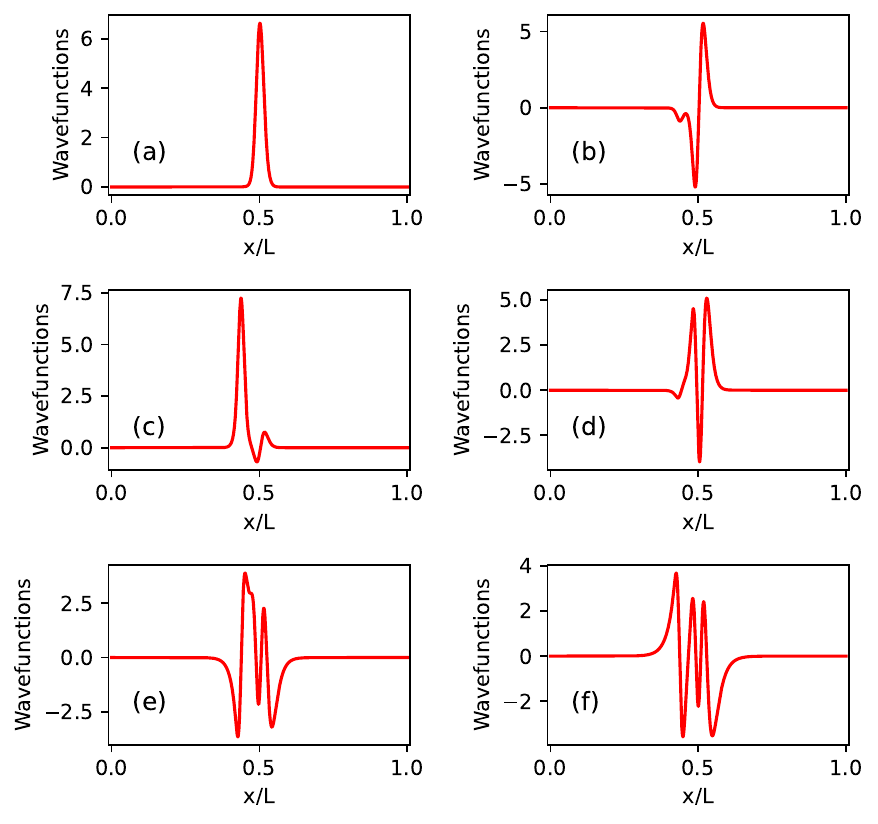}
\caption{Lowest six bound states of the Schr\"odinger equation within the two outer-wells of the Rb(40S) Rydberg potential. The magnitude of binding energies decreases monotonically from (a) to (f). Panels (a), (b), (d) show the first three bound states in the furthest well, panel (c) depicts the lowest bound state in the secondary well, and panels (e) and (f) show the bound states occupying both wells. The furthest Rydberg-potential well is located at $x=L/2$.
}
\label{Fig:Rbn62_SP}
\end{figure}

\section{Rydberg molecules in noninteracting gases}
The energetically lowest six bound states obtained from the Schr\"odinger equation with the outer double-well approximation of the Rb(40S) Rydberg potential are provided in Fig.~\ref{Fig:Rbn62_SP}. Here, we follow the same scaling process (i.e., $\alpha=0.5$) to place the furthest dip of the Rydberg potential at the middle of the box. We consider only one component of fermions here, as the other component will result in exactly the same results in a noninteracting system. By analyzing the number of nodes and the localization position of the wave functions, one can identify the series of the bound states localized in the furthest well of the Rydberg potential and the ones located in the inner well. For example, panels (a), (b), (d), (e) of Fig.~\ref{Fig:Rbn62_SP} showcase the first four bound states localized in the furthest dip, while panels (c) and (f) depict the two lowest bound states localized in the secondary well. 
Given the well-separated energies of the bound states, all bound states shown here correspond to diatomic Rydberg molecules between the Rydberg atom and a fermion. Their binding energies are extracted from the eigenvalues of the Schr\"odinger equation with the Rydberg potential. After proper normalization with respect to the values of $E_f$ for the corresponding BCS and BEC cases, the binding energies are shown in Fig.~\ref{Fig:Eb_Rbn62} of the main text. We note in passing that triatomic Rydberg molecules binding a noninteracting spin-$\uparrow$ fermion and a noninteracting spin-$\downarrow$ fermion can also form. However, they will possess twice the binding energy of a corresponding diatomic Rydberg molecule. This is because for noninteracting fermions, each energy level can be occupied by one spin-$\uparrow$ and one spin-$\downarrow$ fermion since they do not feel the presence of each other. It should be emphasized that this binding process is different from the one of the triatomic Rydberg molecule encompassing a trapped Cooper pair, where pairing effects play a decisive role.

\end{document}